\documentclass[proof]{pasj00}
\begin{document}
\SetRunningHead{T. Nakajima et al.}{Multi-Transition Study of Molecules in NGC 1068}
%\Received{}%{yyyy/mm/dd}
%\Accepted{}%{yyyy/mm/dd}
%\Published{}%{yyyy/mm/dd}

\title{A Multi-Transition Study of Molecules toward NGC 1068 based on High-Resolution Imaging Observations with ALMA}

\author{%
  Taku \textsc{Nakajima},\altaffilmark{1}
  Shuro \textsc{Takano},\altaffilmark{2,3}
  Kotaro \textsc{Kohno},\altaffilmark{4,5}
  Nanase \textsc{Harada},\altaffilmark{6}
  Eric \textsc{Herbst},\altaffilmark{7}\\
  Yoichi \textsc{Tamura},\altaffilmark{4}
  Takuma \textsc{Izumi},\altaffilmark{4}
  Akio \textsc{Taniguchi},\altaffilmark{4}
  and
  Tomoka \textsc{Tosaki}\altaffilmark{8}
}
\altaffiltext{1}{Solar-Terrestrial Environment Laboratory, Nagoya University, \\Furo-cho, Chikusa-ku, Nagoya, Aichi, 464-8601}
\email{nakajima@stelab.nagoya-u.ac.jp}
\altaffiltext{2}{Nobeyama Radio Observatory, National Astronomical Observatory of Japan, \\462-2, Nobeyama, Minamimaki, Minamisaku, Nagano, 384-1305}
\altaffiltext{3}{Department of Astronomical Science, The Graduate University for Advanced Studies (SOKENDAI), \\462-2, Nobeyama, Minamimaki, Minamisaku, Nagano, 384-1305}
\altaffiltext{4}{Institute of Astronomy, The University of Tokyo, \\2-21-1, Osawa, Mitaka, Tokyo, 181-0015}
\altaffiltext{5}{Research Center for Early Universe, School of Science, The University of Tokyo, \\7-3-1, Hongo, Bunkyo-ku, Tokyo, 113-0033}
\altaffiltext{6}{Max Planck Institute for Radio Astronomy, \\D-53121 Bonn, Germany}
\altaffiltext{7}{Department of Chemistry, University of Virginia, McCormick Road, PO Box 400319, \\Charlottesville, VA 22904, USA}
\altaffiltext{8}{Department of Geoscience, Joetsu University of Education, \\Yamayashiki-machi, Joetsu, Niigata, 943-8512}


\maketitle

\begin{abstract}
We present 0.8-mm band molecular images and spectra obtained with the Atacama Large Millimeter/submillimeter Array (ALMA) toward one of the nearest galaxies with an active galactic nucleus (AGN), NGC 1068. Distributions of CO isotopic species ($^{13}$CO and C$^{18}$O) $\it{J}$ = 3--2, CN $\it{N}$ = 3--2 and CS $\it{J}$ = 7--6 are observed toward the circumnuclear disk (CND) and a part of the starburst ring with an angular resolution of $\sim$1.$\!^{\prime\prime}$3 $\times$ 1.$\!^{\prime\prime}$2. The physical properties of these molecules and shock-related molecules such as HNCO, CH$_{3}$CN, SO, and CH$_{3}$OH detected in the 3-mm band were estimated using rotation diagrams under the assumption of local thermodynamic equilibrium. The rotational temperatures of the CO isotopic species and the shock-related molecules in the CND are, respectively, 14--22 K and upper limits of 20--40 K. Although the column densities of the CO isotopic species in the CND are only from one-fifth to one-third of that in the starburst ring, those of the shock-related molecules are enhanced by a factor of 3--10 in the CND. We also discuss the chemistry of each species, and compare the fractional abundances in the CND and starburst ring with those of Galactic sources such as cold cores, hot cores, and shocked molecular clouds in order to study the overall characteristics. We find that the abundances of shock-related molecules are more similar to abundances in hot cores and/or shocked clouds than to cold cores. The CND hosts relatively complex molecules, which are often associated with shocked molecular clouds or hot cores. Because a high X-ray flux can dissociate these molecules, they must also reside in regions shielded from X-rays.
\end{abstract}

\section{Introduction}

It is important to investigate the relationships between the power sources of galaxies (i.e., active galactic nuclei (AGN) and/or starbursts) and the chemical properties of the surrounding dense interstellar medium in order to study the effects on molecular abundances and to probe the power source with molecular line observations. Chemical properties, especially, have been expected to be powerful astrophysical tools for the study of galaxies, because the molecular line observations of different galaxies allow us to study the effects of these different physical properties/activities on the molecular medium. In fact, some groups have suggested that it is possible to diagnose power sources in dusty galaxies using molecular line ratios in nearby galaxies (e.g., Kohno et al. 2001, 2008; Usero et al. 2004; Kohno 2005; Imanishi et al. 2007; Krips et al. 2008; Izumi et al. 2013). For example, elevated HCN emission with respect to CO and/or HCO$^{+}$ has often been detected toward AGNs (e.g., Kohno et al. 1996, 2003), where it is expected to be the imprint of either strong X-ray irradiation/ionization (e.g. Garc\'{i}a-Burillo et al. 2010) and/or a high-temperature environment caused by AGN activity (e.g., Harada et al. 2010; Izumi et al. 2013). On the other hand, we reported no significant differences in the relative abundances of the carbon containing molecules C$_{2}$H and cyclic-C$_{3}$H$_{2}$ between one of the nearest galaxies with an AGN, NGC 1068, and the prototypical starburst galaxy, NGC 253. It was concluded that these basic carbon-containing molecules are insensitive to AGNs and/or these molecules exist in a cold gas away from an AGN (Nakajima et al. 2011).

Systematic unbiased scans (i.e. molecular line survey observations) are the most effective method not only for complete understanding of chemical compositions in representative sources, but also for probing interstellar medium and star formation. Over the last 10 years or so, millimeter/sub-millimeter observing systems with very high sensitivity, wide frequency range, and high velocity resolution have been put to use in a number of telescopes (e.g. Carter et al. 2012; Nakajima et al. 2008; 2013a). With these, we can also obtain higher quality line survey observations of millimeter/sub-millimeter molecular lines toward nearby galaxies as well as Galactic sources. To date, several line surveys have been reported towards the center of NGC 1068 using single dish telescopes (Snell et al. 2011; Costagliola et al. 2011; Kamenetzky et al. 2011; Spinoglio et al. 2012; Aladro et al. 2013; Nakajima et al. 2013b). For example, Aladro et al. (2013) compared their results with NGC 253, and they suggested that NGC 1068 has a different chemical composition from those of starburst galaxies. They discussed that CH$_{3}$CCH is not detected in NGC 1068, yet CN, SiO, HCO$^{+}$, and HCN are detected and enhanced. Although the chemical environment of NGC 1068 has become clearer thanks to these line survey observations, contamination from starbursts associated with inner spiral arms/ring ($d\sim$30$^{\prime\prime}$) could be a problem, if we consider the sizes of the observing beams (15$^{\prime\prime}$--70$^{\prime\prime}$) in these single dish telescopes. On the other hand, interferometric imaging of the circumnuclear disk (CND) gives clean measurements of spectral lines, but the observed lines are limited to major species such as $^{12}$CO ($\it{J}$ = 1--0, 2--1 and 3--2), $^{13}$CO ($\it{J}$ = 1--0, 2--1 and 3--2), C$^{18}$O ($\it{J}$ = 1--0 and 2--1), HCN ($\it{J}$ = 1--0 and 3--2), HCO$^{+}$ ($\it{J}$ = 1--0, 3--2 and 4--3), CS ($\it{J}$ = 2--1), CN ($\it{N}$ = 2--1), and SiO ($\it{J}$ = 2--1) (e.g. Planesas et al. 1991; Kaneko et al. 1992; Jackson et al. 1993; Tacconi et al. 1994, 1997; Helfer and Blitz 1995; Papadopoulos et al. 1996; Schinnerer et al. 2000; Kohno et al. 2008; Garc\'{i}a-Burillo et al. 2010; Krips et al. 2011; Tsai et al. 2012) due to the limitation in sensitivity of the existing pre-ALMA (Atacama Large Millimeter/sub-millimeter Array) interferometers. Although it is important to estimate the physical properties, such as temperature and density and so on, only one transition has been observed except for CO, HCN and HCO$^{+}$.

In this situation, we proposed to observe several interesting molecules sensitively with ALMA in the 96--100 GHz and 108--111 GHz (3-mm) regions for lower excitation lines (Takano et al. 2014, hereafter Paper-1) and also 327.5--330.5 GHz and 338.5--342.5 GHz (0.8-mm) regions for higher excitation lines. These frequency regions are rich in molecules, including typical shock/dust related species and the CO isotopologues. Our aim has been to focus on the effects of strong X-ray and starburst on shock and/or dust related molecules, CH$_{3}$OH, SO, HNCO, and CH$_{3}$CN, which are well observed in Galactic sources to study chemical and physical conditions. Our observational results with ALMA will be compared with observations in Galactic sources, model calculations, and laboratory experiments to study formation and destruction mechanisms of the above molecules. The complex organic molecules CH$_{3}$OH and CH$_{3}$CN are thought to be efficiently produced on dust (e.g. Watanabe and Kouchi 2002, Takano et al. 1995) and subsequently desorbed into gas-phase (Garrod et al. 2008), while HNCO is also thought to be produced at least partially on dust (Quan et al. 2010). The radical SO is thought to have an enhanced abundance in shocked regions (Leen and Graff 1988). However, the abundances and chemical processes forming and destroying these molecules are not well understood in galaxies with active galactic nuclei.

From our observations in the two frequency bands with ALMA, rotational temperatures and column densities can be obtained using intensities of two transitions. The field of view of the 0.8-mm observations is about 18$^{\prime\prime}$, which means that only the CND region can be covered with one pointing in NGC 1068. Therefore, we additionally observe a part of the starburst ring at the offset position from the center, so that distributions and abundances of molecules in the starburst environment as well as the CND can be studied. This comparison will enable us to further study the impact of AGNs on the surrounding dense interstellar medium. In Paper-1, we have presented the distributions of detected molecules in the 3-mm band, such as $^{13}$CO, C$^{18}$O, $^{13}$CN, CS, SO, HNCO, HC$_{3}$N, CH$_{3}$OH, CH$_{3}$CN, and discussed the implications of diversity. The main results of Paper-1 are that the molecular distributions are reflections of both physical and chemical properties.

In this paper, we report a high-resolution imaging study of molecular lines in the 0.8-mm band toward NGC 1068 observed with ALMA, and also the calculation of the fractional abundances of shock/dust related molecules based on not only the observations in 0.8-mm band but also on the earlier 3-mm band results from Paper-1. Even in its early science operation phase, ALMA was already powerful enough to simultaneously observe four high-excitation molecular lines ($^{13}$CO and C$^{18}$O $\it{J}$ = 3--2, CN $\it{N}$ = 3--2 and CS $\it{J}$ = 7--6) in the 0.8-mm band, uncovering a wide variety of molecular line distributions. We describe our observations and data reduction in Section 2, and present the images and spectra of the observed molecular lines and also estimated physical properties of individual molecules in Section 3. In Section 4, we discuss relationships among the molecular abundances toward the CND and starburst ring in NGC 1068, and Galactic sources such as hot cores, cold cores, and shocked clouds. Throughout the paper, we assume that the distance of NGC 1068 is 14.4 Mpc (Tully 1988; Bland-Hawthorn et al. 1997); at this distance, 1$^{\prime\prime}$ corresponds to 72 pc.

\section{Observations}

We observed NGC 1068 using ALMA in Cycle 0 (Early Science program) with band-3 ($\lambda\sim$3 mm) and band-7 ($\lambda\sim$0.8 mm) receivers in January of 2012 and November of 2011, respectively (ID = 2011.0.00061.S; PI = S. Takano). The details of the 3-mm band observations are described in Paper-1, and the 0.8-mm band observations are explained below. Table 1 summarizes the observational parameters of the 0.8-mm band. The centers of four spectral windows were tuned to 329.50 GHz (spw0), 330.85 GHz (spw1), 340.90 GHz (spw2), and 342.80 GHz (spw3), each with a 1.875 GHz bandwidth and 3840 channels of 488.28 kHz width. The final results were presented with a velocity resolution of $\sim$20 km s$^{-1}$ to improve the signal-to-noise ratio. The 14 operational 12-m antennas in the compact array configuration were arranged to span baseline lengths of 20--196 m, which gave a beam size of 1.$\!^{\prime\prime}$3 $\times$ 1.$\!^{\prime\prime}$2 ($\sim$90 $\times$ 86 pc) for the lowest frequency, spw0. A two-pointing mosaic was observed in order to image the CND of $\lesssim$280 pc ($\lesssim$4$^{\prime\prime}$) in diameter and the surrounding starburst ring of $\sim$2 kpc ($\sim$30$^{\prime\prime}$) in diameter (Schinnerer et al. 2000). The phase reference center has been set to $\alpha_{J2000}$ = 2$^{h}$42$^{m}$40.$\!^{s}$798 and $\delta_{J2000}$ = -00$^{\circ}$00$\!^{\prime}$47$\!^{\prime\prime}$938 (Schinnerer et al. 2000), which corresponds to the radio position of the active nucleus (Muxlow et al. 1996). The position of a part of the starburst ring is at the offset position to the southwest from the AGN of ($\Delta\alpha$, $\Delta\delta$) = (-10$^{\prime\prime}$, -10$^{\prime\prime}$), which is stronger in CO, HCN (Schinnerer et al. 2000, Kohno et al. 2008). The systemic velocity employed was 1150 km s$^{-1}$, which was taken from Schinnerer et al. (2000). The total integration times on the CND and starburst ring were approximately 30.3 minutes in each position and the total observation time was 71.4 minutes. The system noise temperatures including the atmosphere were $\sim$120--250 K and $\sim$150--700 K in the upper sideband (USB) and the lower sideband (LSB), respectively, depending on frequency and antennas. The delivered data were already bandpass, flux, and gain (phase and amplitude) calibrated by the ALMA Regional Center. All of the above calibration and imaging were carried out using the Common Astronomy Software Applications (CASA; McMullin et al. 2007). 

\section{Results}

\subsection{Molecular gas distributions}

We have successfully detected C$^{18}$O ($\it{J}$ = 3--2), $^{13}$CO ($\it{J}$ = 3--2), CN ($\it{N}$ = 3--2) and CS ($\it{J}$ = 7--6) in the CND, and the same molecules except for CS were also detected in the starburst ring. Figure 1 shows the integrated intensity maps of four clearly detected molecules. These are two-point mosaic images toward the CND and a part of the starburst ring. The C$^{18}$O emission was not significantly detected toward the CND, but relatively strong in the starburst ring. We found a similar feature for C$^{18}$O ($\it{J}$ = 1--0) in the 3-mm band observations (Paper-1). On the other hand, $^{13}$CO ($\it{J}$ = 3--2) is very strong in the CND and also detected towards the starburst ring. The CN and CS radicals are very weak and not detected towards the starburst ring, respectively, but both are clearly detected towards the CND (see also the spectra in Fig.2). Moreover, the $^{13}$CO, and CN emission features from the CND are spatially separated into two conspicuous knots, and the CS emission is also separated into two knots, albeit faintly.

The difference in the C$^{18}$O and $^{13}$CO distributions cannot be explained by different excitation conditions, because the abundance ratio of these species is not sensitive to the molecular gas temperature and density. Therefore, this feature is possibly the result of differences in isotopic ratios. The [$^{12}$C]/[$^{13}$C] isotopic ratios in the Galaxy can be quite diverse. For example,  the ratio is 20--25 in the Galactic center clouds (G\"{u}sten et al. 1985); $\sim$89 in the solar system (Wilson \& Matteucci 1992) and outer galaxies (e.g. $\sim$40 (Henkel et al. 1993) and $\sim$80 (Mart\'{i}n et al. 2010) for NGC 253; and $>$100 for the Clover leaf quasar (Henkel et al. 2010)). Given this diversity, differences in the isotopic ratios  [$^{12}$C]/[$^{13}$C] and [$^{16}$O]/[$^{18}$O] between the CND and the starburst ring of NGC 1068 are possible. Another possibility is the effect of isotope-selective photodissociation, in which the abundance of naturally poor isotopic species are decreased because of their lower optical depth (e.g. Harrison et al. 1995). Details concerning these discussions for NGC 1068 will appear in a separate paper (Taniguchi et al., in preparation). 

\subsection{Line properties in the CND and starburst ring}

Figure 2 shows the spectra of all spectral windows toward the CND and starburst ring at the southwest position, which is the $^{13}$CO ($\it{J}$ = 3--2) intensity peak, as mentioned in the caption of Figure 2. Table 2 and 3 list the Gaussian fitted line parameters toward the CND and starburst ring, respectively. The spectra of the LSB consist of spw0 (327.29--330.16 GHz) and spw1 (328.49--330.49 GHz), those of the USB consist of spw2 (338.68--340.54 GHz) and spw3 (340.56--342.43 GHz). To obtain the spectra and measure the Gaussian fitted line parameters, we convolved all 0.8-mm band images to the beam size of the 3-mm band data (4.$\!^{\prime\prime}$21$\times$2.$\!^{\prime\prime}$36, P.A. = 176$^{\circ}$), and then the correction of the primary beam attenuation was applied. $^{13}$CO, CN and CS were clearly detected toward the CND although C$^{18}$O was not clearly detected ($\sim$4 $\sigma$). On the other hand, C$^{18}$O, $^{13}$CO and CN were clearly detected toward the starburst ring. Unfortunately, expected lines of shock and/or dust related molecules, CH$_{3}$OH, SO, HNCO, and CH$_{3}$CN, were not detected. We calculated the upper limits of the integrated flux (1$\sigma$) for these molecules from the rms noise level. The line widths are broad in the CND with approximately 170--180 km s$^{-1}$ except for CN. For CN, two partially separated fine structure components of $\it{J}$ = 5/2--3/2 and $\it{J}$ = 7/2--5/2 were identified at the band edge. Therefore the line parameters of CN may not be very precise. Moreover, it is not possible to resolve each hyperfine component. Unlike the case with the CND, the line widths of all detected lines in the starburst ring are narrow, approximately 30--40 km s$^{-1}$. 

\subsection{Comparison of flux with single dish telescope}

We compared the flux obtained with ALMA with those obtained with the single dish James Clerk Maxwell Telescope (JCMT) to estimate the recovery of the flux with the interferometer. For $^{13}$CO $\it{J}$ = 3--2, the image obtained with ALMA (primary beam corrected) was convolved with the JCMT beam of 14$^{\prime\prime}$, and the flux obtained was converted to the brightness temperature, which is 3.4$\pm$0.1 K km s$^{-1}$. The corresponding value obtained with the JCMT is 10.7$\pm$1.3 K km s$^{-1}$, which is observed by Israel (2009). Therefore, the ALMA observations recover 32$\pm$5 \% of the single dish flux. For CS $\it{J}$ = 7--6, the flux obtained with ALMA was converted to brightness temperature, which is $\sim$0.4 K km s$^{-1}$. Although the detection of CS $\it{J}$ = 7--6 with the JCMT is marginal, the corresponding value is $\sim$1.4 K km s$^{-1}$ (Bayet et al. 2009). Therefore, the recovered flux is about 26 \%. The recovered flux of $^{13}$CO $\it{J}$ = 1--0 in the 3-mm band is about 80--90 \%, which was obtained by comparing with the single dish telescopes NRO 45-m and IRAM 30-m (see the section 3.4 in Paper-1). The probably reason for the increase in the missing flux in the 0.8-mm band is that the effective minimum baseline of the observations in this band is longer than that for the observations in the 3-mm band

\subsection{LTE analysis of individual molecules}
We constructed the rotation diagrams of observed molecules in the 3-mm band (Paper-1) and the 0.8-mm band (this work) toward each position of the CND and starburst ring (figure 3). All 0.8-mm band images were convolved to the beam size of the 3-mm band data. Thus, the data points of the 3-mm and 0.8-mm bands were obtained from the emission in the same regions. The rotational temperatures ($T_{\rm{rot}}$) and column densities ($N_{\rm{mol}}$) are calculated from the slope and intercept at $E_{u}$ = 0 of the diagram under the assumptions that all lines are optically thin, and that a single excitation temperature ($T_{\rm{ex}}$) characterizes all transitions (the local thermodynamic equilibrium (LTE) assumption) (Turner 1991). The non-detected lines are plotted with their rms noise level (3$\sigma$), and the upper or lower limits of physical parameters are calculated. For the calculation of $N_{\rm{mol}}$ for molecules with only one detected transition, we assumed some temperatures of $T_{\rm{rot}}$ = 5 K, 10 K, 15 K, 30 K and 45 K. The results of all molecules are shown in Tables 4 and 5. The calculated rotation temperatures and column densities of the observed molecules, C$^{18}$O in the starburst ring, CS in the CND, and $^{13}$CO and HC$_{3}$N in both regions are listed in Table 4. The column densities for the assumed some rotation temperatures of C$^{18}$O and CH$_{3}$CN (CND), HNCO, CN, SO, and CH$_{3}$OH (CND and starburst ring), and CS (starburst ring) are listed in Table 5, because these molecules were detected in only one transition. Note that these rotation diagrams are fitted with only two points, so that the possibility of multiple components at different temperatures cannot be ruled out (Bayet et al. 2009).

\subsubsection{C$^{18}$O and $^{13}$CO}
The rotation diagrams of C$^{18}$O and $^{13}$CO are shown in Figure 3 (a) and (c), respectively. The C$^{18}$O $\it{J}$ = 1--0 line was not detected toward the CND, thus we obtained a lower limit of $T_{\rm{rot}}$ $>$ 14.2 K. These CO isotopic molecules trace cold gas (8.0 K and 8.5 K) in the starburst ring and some warm gas ($>$ 14.2 K, 21.5 K) in the CND. Aladro et al. (2013) suggested that $^{13}$CO has two different components in the rotation diagram, 4.3 K and 24.1 K. In that work, it is possible that the reported emissions from the CND and starburst ring are blended, because they used a single dish telescope. In addition, they obtained that the $T_{\rm{rot}}$ of C$^{18}$O is the lowest in all observed molecules, 3.3 K. Therefore, as compared with our result, previous work with the single dish telescope overestimated the intensity of a vertical axis in the rotation diagram for the purpose of examination of property in the CND. Or, as we saw in Section 3.3, the flux of the $\it{J}$ = 3--2 line in the starburst ring with ALMA is not sufficiently recovered for this study. If the previous results with the single dish telescope are accurate enough, the flux in the starburst ring with ALMA should be increased by a factor of 4.5. Then the $T_{\rm{rot}}$ and column density of $^{13}$CO will be 16.3 K and 7.1$\times$10$^{16}$ cm$^{-2}$, respectively. We leave the effect of missing flux out of consideration, because these values make little impact on the following discussion.

Unlike other molecules, the column densities of both CO isotopologs in the starburst ring are larger than those in the CND. The column density of $^{13}$CO is a factor of 3.5, and C$^{18}$O a factor of 2.4 larger than those in the CND under the assumption of $T_{\rm{rot}}$ = 10 K. This tendency is the same with $^{12}$CO. Tsai et al. (2012) calculated the $N_{\rm H_{2}}$ in the CND and the starburst ring from the integrated intensity of $^{12}$CO $\it{J}$ = 1--0 line. $N_{\rm H_{2}}$ in the starburst ring at the area R15, which includes the area of our observations at the southwest region in the starburst ring, is a factor of about 20 larger than those in the CND. On the other hand, although the $^{13}$CO $\it{J}$ = 1--0 line is rather weak and C$^{18}$O $\it{J}$ = 1--0 line is not detected toward the CND (Paper-1), emission of high-$J$ transition lines such as $^{13}$CO $\it{J}$ = 3--2 in our observation and $^{12}$CO $\it{J}$ = 6--5 based on the data by Garc\'{i}a-Burillo et al. (2014) are strong and clearly detected. Thus, the kinetic temperature in the CND is very high and presumably in a non-LTE environment. The details of the physical properties based on a large velocity gradiant (LVG) analysis of the CO lines will be described in Taniguchi et al. (in preparation).

\subsubsection{HNCO and CH$_{3}$OH}
The rotation diagrams of HNCO and CH$_{3}$OH are shown in Figure 3 (b) and (g), respectively. The HNCO $\it{J_{Ka,Kc}}$ = 15$_{0,15}$--14$_{0,14}$ and CH$_{3}$OH $\it{J_{Ka,Kc}}$ = 13$_{1,12}$--13$_{0,13}A$-+ lines were not detected toward both the CND and the starburst ring, thus we could obtain only upper limits of $T_{\rm{rot}}$. The CH$_{3}$OH $\it{J_{K}}$ = 2$_{K}$--1$_{K}$ lines are composed of four transitions (2$_{-1}$--1$_{-2}E$, 2$_{0}$--1$_{0}A$+, 2$_{0}$--1$_{0}E$, 2$_{1}$--1$_{1}E$), but the quartet of lines is blended in the spectra from NGC 1068. Therefore, we calculated $T_{\rm{rot}}$ using the method for blended lines by Mart\'{i}n et al. (2006). $T_{\rm{rot}}$ in the CND and the starburst ring is less than 28.6 K and 36.0 K, respectively, for HNCO, and 40.4 K and 38.9 K, respectively, for CH$_{3}$OH. It is possible that these lines trace warm gas ($\lesssim$ 30--40 K). This value of HNCO is consistent with the value of $\sim$30 K obtained by Aladro et al. (2013). However, they obtained 5.6 K for CH$_{3}$OH, which is almost six times lower than our result. The column densities of HNCO in the CND are a factor of $\sim$5 larger than those in the starburst ring assuming a $T_{\rm{rot}}$ of 10--30 K. On the other hand, the column density of CH$_{3}$OH is not much different between the CND and the starburst ring under the assumption of a similar $T_{\rm{rot}}$. If the $T_{\rm{rot}}$ in the CND is higher than the starburst ring (e.g. 30 K in the CND and 10 K in the starburst ring), the column density of CH$_{3}$OH in the CND is a factor of $\sim$3 larger than that in the starburst ring.

\subsubsection{CH$_{3}$CN}
The rotation diagram is shown in Figure 3 (d). The $\it{J_{K}}$=6$_{K}$--5$_{K}$ line toward the starburst ring and the $\it{J_{K}}$=18$_{K}$--17$_{K}$ lines toward both positions are not detected. We obtained an upper limit of $T_{\rm{rot}}$, $<$ 26.1 K in the CND. The column density in the CND ($\sim$1$\times$10$^{13}$ cm$^{-2}$) is the smallest among the observed shock/dust related molecules. CH$_{3}$CN is a tracer of star formation in Galactic sources, such as hot cores and outflows from young stellar objects. Therefore, the non-detection of this molecule in the starburst ring, but the detection in the CND are interesting results. We discuss the abundance of this molecule in Section 4.2.2.

\subsubsection{CN}
The rotation diagram is shown in Figure 3 (e). We did not observe CN in the 3-mm band, thus we plotted only the $\it{N}$ = 3--2 transition in the 0.8-mm band. In line survey observations with the NRO 45-m telescope, we found that the CN/$^{13}$CO intensity ratios are significantly higher in NGC 1068 ($\sim$2.3) than in NGC 253 ($\sim$1.1) and IC 342 ($\sim$0.6) (Nakajima et al. 2013b). The CN/$^{13}$CO intensity ratios are $\sim$4.3 in the CND and $\sim$0.1 in the starburst ring with this ALMA observation. Moreover, the column density of CN in the CND is a factor of 20--50 larger than that in the starburst ring. CN is one of the key molecules to trace X-ray dominated regions (XDR), as mentioned in theoretical work (e.g. Meijerink et al. 2007). Therefore, the enhancement of intensity and column density of CN in the CND may be the effects of extreme physical conditions such as those that pertain to XDRs.

\subsubsection{SO}
The rotation diagram is shown in Figure 3 (f). The $\it{J_{N}}$=7$_{8}$--6$_{7}$ line is not detected toward both the CND and the starburst ring. We obtained upper limits of $T_{\rm{rot}}$ in the CND and the starburst ring of 20--26 K. This temperature is consistent with the value of 22.8 K of Aladro et al. (2013). The previous work of SO, with a single dish telescope, traced mainly the CND, because the distribution of SO is concentrated in the CND and the contribution is less from the starburst ring, as can be seen in Figure 2 in Paper-1. The column density in the CND is a factor of 5--6 larger than that in the starburst ring assuming the same $T_{\rm{rot}}$. If the rotational temperature in the CND is higher than in the starburst ring, the ratio of column densities changes. For example, if the rotational temperature in the CND is 30 K and that in the starburst ring 10 K, the column densities will be 2.4 $\times$ 10$^{14}$ cm$^{-2}$ and 0.2 $\times$ 10$^{14}$ cm$^{-2}$, respectively, leading to a column density of approximately 12.

\subsubsection{CS}
The rotation diagram is shown in Figure 3 (h). The $\it{J}$=7--6 line is not detected toward the starburst ring. We obtained an upper limit of $T_{\rm{rot}}$, $<$ 12.1 K, in this region. $T_{\rm{rot}}$ in the CND is 13.1 K, which is consistent with the result of 13.9 K by Bayet et al. (2009). Since the $\it{J}$ = 7--6 line was only marginally detected by Bayet et al.(2009), they obtained $T_{\rm{rot}}$ = 7.1 K from a rotation diagram excluding this line. However, our results support the value including the $\it{J}$ = 7--6 line. Moreover, Bayet et al. (2009) suggested two fitting components, but the CS molecule is distributed not only in the CND, but also in the starburst ring (see figure 1 in Paper-1). Thus, it is possible to have a blending of emission from the starburst ring.

\subsubsection{HC$_{3}$N}
The rotation diagram is shown in Figure 3 (i). We did not observe HC$_{3}$N in the 0.8-mm band, and therefore we obtained only the $\it{J}$ = 11--10 and 12--11 lines in the 3-mm band. Using these lines, $T_{\rm{rot}}$ in the CND and the starburst ring are 22.1 K and 17.7 K, respectively. These values are not consistent with that of Aladro et al. (2013), where they obtained 7.3 K. The reason for the discrepancy is not clear, but, as can be seen, the value for $\it{J}$ = 10--9 is afield from the fitting line of the values for $\it{J}$ = 11--10 and 12--11 in their rotation diagram. Therefore, HC$_{3}$N has possibly more than one component in the CND. The column density of HC$_{3}$N in the CND is about an order of magnitude larger than that in the starburst ring.

\section{Discussion}

\subsection{Knot components in the CND}

The asymmetrical distributions of $^{13}$CO, CN and CS in the CND can be seen in Figure 1. Usero et al. (2004) suggested that two velocity components, which are velocity $<$ systemic velocity and velocity $>$ systemic velocity, correspond to the Eastern knot (E-knot) and Western knot (W-knot), respectively, in the spectra of CO, CS, HCN, SiO, H$^{13}$CO$^{+}$, HCO$^{+}$, and HOC$^{+}$. The E-knot and W-knot are clearly separated in the integrated intensity maps of $^{13}$CO, CN, and CS (Figure 1) in our observations. So far, $^{12}$CO $\it{J}$=1--0 and 2--1 (Schinnerrer et al. 2000), CN $\it{N}$=2--1 (Garc\'{i}a-Burillo et al. 2010), HCN $\it{J}$=3--2, HCO$^{+}$ $\it{J}$=3--2, $^{12}$CO $\it{J}$=3--2, and $^{13}$CO $\it{J}$=1--0, and 2--1 (Krips et al. 2011) have been found to be associated with both the E-knot and the W-knot with high resolution interferometric observations. We find similar distributions for $^{13}$CO ($\it{J}$=3--2), CN ($\it{N}$=3--2) and CS ($\it{J}$=7--6) for the first time with our ALMA observations. 

We have measured the peak positions for the E-knot and W-knot seen in these lines, as well as the flux ratios of the E-knot/W-knot for each molecule, both of which are listed in Table 6. All measured peak positions of the E-knot and W-knot seen in these molecules are consistent with each other. The flux ratios between the E-knot and W-knot are 2.2$\pm$0.2 for $^{13} $CO, 2.3$\pm$0.4 for CN and 3.5$\pm$0.7 for CS; i.e., the E-knot/W-knot flux ratio of CS is higher than those of $^{13}$CO and CN. Krips et al. (2011) obtained the E-knot/W-knot flux ratios of $^{12}$CO ($\it{J}$=1--0, 2--1 and 3--2), $^{13}$CO ($\it{J}$=1--0 and 2--1), HCN ($\it{J}$=3--2) and HCO$^{+}$ ($\it{J}$=3--2) to be 1.0--2.3 with SMA and PdBI observations, and Usero et al. (2004) shows the ratios of $^{12}$CO ($\it{J}$=1--0 and 2--1), HCN ($\it{J}$=1--0), SiO ($\it{J}$=2--1 and 3--2), H$^{13}$CO$^{+}$ ($\it{J}$=1--0), HCO$^{+}$ ($\it{J}$=1--0), and HOC$^{+}$ ($\it{J}$=1--0) with IRAM 30 -m observations to be 0.6--1.8 and CS ($\it{J}$=2--1) to be 2.2. Therefore, the measured E-knot/W-knot flux ratio of CS ($\it{J}$=7--6) is higher than any other molecular lines. These results suggest that there is a chemical differentiation between the E and W knots in the CND (e.g. Usero et al. 2004). However, the details of the chemical differentiation and/or the physical structure in the CND are still not clear based on our observations with the spatial resolution available with ALMA during Cycle 0.

\subsection{Fractional Abundances}
In order to understand which mechanism, such as X-rays, shock waves, or ultraviolet photons, dominates the observed chemical features, it is helpful to obtain fractional abundances of each species with respect to molecular hydrogen. There are several different possible driving forces of chemistry in the observed regions of NGC 1068. For example, since the AGN emits intense X-rays, it is likely to have XDRs (e.g., Garc\'{i}a-Burillo et al., 2010). Elevated massive star formation can create photon-dominated regions (PDRs; e.g., Hollenbach \& Tielens 1999) due to ultraviolet radiation from OB stars. In the site of embedded on-going star formation, hot cores can also contribute to molecular emission. Furthermore, the shock waves affect the chemistry by heating up the gas as well as by sputtering dust grains. With our beam size ($\sim$90 pc), a mixture of molecular clouds governed by these different mechanisms may be observed. Less dense gas than the mean density of the molecular cloud is likely to have a higher volume-filling factor, while emission from compact sources such as hot cores tends to have much smaller volume-filling factors unless the region hosts an extremely large number of hot cores. With this complexity in mind, we derived the fractional abundances and compared our results with various types of Galactic sources to understand the nature of molecular material in the CND and starburst ring in NGC 1068.

To compute fractional abundances, we first need to estimate the column density of molecular hydrogen $N_{\rm H_2}$, which is often derived from observations of CO molecules. Here we adopt $N_{\rm H_2}$ values of $7.4 \times 10^{21}$ cm$^{-2}$ at the CND and $2.6 \times 10^{22}$ cm$^{-2}$ at the starburst ring, based on the estimated column densities of $^{13}$CO, which is an optically thin tracer of molecular gas mass. We computed these by the rotation diagram analysis of our ALMA data (summarized in Table 4) with an assumption of a [$^{13}$CO]/[H$_{2}$] fractional abundance of $2 \times 10^{-6}$ (Dickman 1978). These values can be compared with $N_{\rm H_2}$ values based on $^{12}$CO line measurements adopting a CO-to-H$_{2}$ conversion factor ($X_{\rm CO}$ factor) to check the consistency with previous studies. This approach was taken by Tsai et al. (2013), who derived $N_{\rm H_2}$ values of $5.6 \times 10^{21}$ cm$^{-2}$ at the CND and $1.1 \times 10^{23}$ cm$^{-2}$ at the starburst ring, where a similar angular resolution CO $\it{J}$ = 1--0 image (produced from visibilities with similar minimum $\it{uv}$ distances to our 0.8-mm band data) is available. 
%Note that their $N_{\rm H_2}$ values are based on $^{12}$CO $\it{J}$ = 1--0 line intensities and $X_{\rm CO}$ factors. 
We find that the $N_{\rm H_2}$ values at the CND agree well with each other, whereas we find a factor of $\sim$4 disagreement in the starburst ring between the $N_{\rm H_2}$ values from Tsai et al. (2013) and our $^{13}$CO multi-transition analysis. This disagreement in the starburst region may be caused by uncertainties in the adopted $X_{\rm CO}$ factors (see Bolatto et al. 2013 for a recent review on this issue) and/or in the [$^{13}$CO]/[H$_{2}$] fractional abundance. Therefore, we should bear in mind that the estimated fractional abundances in each molecule may suffer from uncertainties in the $N_{\rm H_2}$ values by a factor of a few at least at the starburst ring.

Table 7 lists the derived fractional abundances of molecules in the CND and starburst ring of NGC 1068. Large differences in the fractional abundances are derived between the two sources for HNCO, CH$_3$CN, CN, SO, CH$_3$OH, CS, and HC$_3$N. There are a number of possible reasons for this difference. First, because the CND is denser than the starburst ring, species with higher critical densities may have higher emission in the CND, as discussed in Paper-1. Also, in the starburst ring, relatively more complex molecules such as HNCO, CH$_{3}$CN, CH$_{3}$OH, and HC$_{3}$N can be more easily photo-dissociated because of the lower density. For granular species, lower abundances in the starburst ring can also result from relatively fewer shocked clouds or hot cores compared with the CND where molecular clouds are heated up so that desorption can occur efficiently, either by gravitational contraction leading to a hot core or by shock waves. Finally, the large uncertainties in the H$_{2}$ column densities can lead to large error bars in the fractional abundances with respect to H$_{2}$. Below we discuss the chemistry of each species, and compare the observed abundances in the CND and starburst ring with those of Galactic sources such as TMC-1 (cold core), Sgr B2(N) and AFGL2591 (hot cores), and L1157 (shocked molecular cloud). Comparisons are mostly done for the CND, because the lower fractional abundances of molecules in the starburst ring can be caused by any reason stated above.

\subsubsection{HNCO}
Quan et al.(2010) show that HNCO can be formed both in the gas phase and on grain surfaces from OCN either through the gas-phase hydrogenation of protonated species or hydrogenation on grains with neutral atomic hydrogen. In cold or lukewarm cores, gas-phase reactions can produce HNCO with fractional abundances with respect to total hydrogen density of about $10^{-10}$, whereas the abundance can rise to $10^{-8}$ when HNCO on the grain surface sublimates in a hot core or its warm envelope. Shock waves are also likely to cause sublimation of HNCO into the gas-phase, and raise the fractional abundance of gaseous HNCO. The average fractional abundance of HNCO in the CND is comparable to that of a hot core or shocked region (Table 7), so that sublimation during warm up or via shock waves in the CND may be causing the high abundance of HNCO there. 

\subsubsection{CH$_{3}$CN}
The peak emission of CH$_3$CN in Paper-1 shows no obvious offset from the central radio continuum position as far as we could tell from the current angular resolution of the map. The formation of CH$_3$CN occurs at lower temperatures through a radiative association reaction in the gas-phase
\begin{equation}
{\rm CH_3^+ + HCN \longrightarrow CH_3CNH^+ }
\end{equation}
followed by 
\begin{equation}
{\rm CH_3CNH^+ + e^- \longrightarrow CH_3CN + H.}
\end{equation}
As the prestellar core condenses isothermally, CH$_3$CN can freeze onto the dust grains. Additional formation routes of CH$_{3}$CN are through surface reactions on dust grains. One of them is a series of hydrogenation reactions starting from C$_{2}$N:
\begin{equation}\label{eq:c2n_hyd}
{\rm C_2N(s) + H(s) \longrightarrow HCCN(s) }
\end{equation}
\begin{equation}
{\rm HCCN(s) + H(s) \longrightarrow CH_2CN(s)}
\end{equation}
\begin{equation}\label{eq:ch2cn_hyd}
{\rm CH_2CN(s) + H(s) \longrightarrow CH_3CN(s),}
\end{equation}
 where (s) stands for solid phase. Another surface reaction to form CH$_{3}$CN is 
 \begin{equation}
{\rm CH_3(s) + CN \longrightarrow CH_3CN(s)}
\end{equation}
 although the route via Reactions (\ref{eq:c2n_hyd})--(\ref{eq:ch2cn_hyd}) is much more efficient in the estimate of the gas-grain model based on the reaction network by Garrod et al. (2008) with some added reactions from Harada et al. (2010). CH$_{3}$CN formed through the above mentioned routes can again desorb into the gas-phase when a protostar is formed and the collapsing gas and dust heats up, forming a hot core. CH$_3$CN is observed with high abundances in hot cores, especially in regions with higher rotational temperatures (Nummelin et al. 2000). Although the overall abundance in the CND is only slightly higher than that of a cold core and equivalent to that of the shocked cloud, the fractional abundance of CH$_3$CN at its peak location might be much higher. Strong emission of CH$_3$CN near the galactic nucleus in the CND might be coming from star formation regions.

\subsubsection{CN}
The distribution of CN emission seems to be clearly separated into the E and W-knots. CN is an unstable radical, and both chemical models and observations suggest that the CN fractional abundance is high when the chemistry is at an early stage of evolution of molecular clouds or when there is high flux of UV-photon, X-rays, and/or cosmic-rays to ionize and dissociate precursor molecules such as HCN. Therefore, its fractional abundance in diffuse clouds is high ($\sim10^{-6}$, in a model by Le Petit et al. 2006), but its fractional abundance decreases as the chemistry evolves to a dense cloud and to hot core/corino. Especially in hot cores, a high-temperature reaction ${\rm CN + H_{2} \longrightarrow HCN + H}$ efficiently converts CN into HCN (Harada et al. 2010). Although this reaction can reduce the fractional abundance of CN at an elevated temperature caused by shock waves, a young shock still has a high fractional abundance of CN (Mitchell \& Deveau, 1983). Even in dense regions, the CN fractional abundance can be abundant in PDRs (Jansen et al. 1995) or XDRs with a fractional abundance of $\sim 10^{-7}$ (Meijerink et al. 2007; Harada et al. 2013). The value derived in the CND is closer to the value in XDRs and in the shocked region L1157(B1), but higher than that of cold cores or, especially, hot cores. Since a high X-ray luminosity from the AGN is seen, the high abundance of CN seen in the CND can be explained by XDRs. Although the emissions of CN peak away from AGNs, a model by Harada et al. (2013) shows that XDRs can exist in relatively less dense regions $\sim$ 100 pc away from the AGN depending on the density structure of the disk. Another possible scenario can be a mixing of ionized/atomic gas with molecular gas caused by turbulence, which is expected to increase the amount of radicals. However, this scenario should be tested by a model.

\subsubsection{CH$_{3}$OH}
As mentioned earlier, the emission of CH$_3$OH is separated into the E-knot and W-knot with weaker emission near the AGN core. The most efficient production mechanism of CH$_3$OH is through grain-surface reactions. A series of hydrogenations starting with CO can occur while the dust is cold ($T \sim 10\,$K). In order for methanol to be observed in the gas-phase, methanol on grains must sublime into the gas-phase. In the cold cores, a small amount of methanol can escape into the gas-phase through non-thermal desorption to make a fractional abundance of about 10$^{-9}$ (Vasyunin \& Herbst 2013). When the dust grains are heated due to either warm-up of the hot core (Garrod et al. 2008) or shock waves (Viti et al. 2011), most of the methanol on grains can sublime into the gas-phase to give a fractional abundance of $\gtrsim 10^{-7} $. The regions that have low abundances of gaseous CH$_3$OH may be free of a mechanism to heat up the dust grains, or the dust can be initially warm so that CO, the precursor species of methanol, desorbs into the gas-phase before it can lead to the formation of methanol. Our observed abundance of CH$_3$OH in the CND is higher than that of cold cores, which suggests that the emission is likely, once again, to be coming from either hot cores or shocked regions. A high abundance of methanol can lead via photolysis and surface chemistry to more complex organic species such as methyl formate (HCOOCH$_{3}$) and dimethyl ether (CH$_{3}$OCH$_{3}$) (Garrod et al. 2008).

\subsubsection{HC$_{3}$N}
The emission of HC$_3$N comes from a compact area near the AGN core (Paper-1), probably corresponding to the E-knot. HC$_3$N can be synthesized in the gas-phase, and the peak fractional abundance in cold cores is 2 $\times 10^{-8}$, a small factor of 1.7 higher than in the CND. The fractional abundance can increase with an elevation of temperature (Harada et al. 2010, 2013), but too high a value for the X-ray ionization rate can dissociate HC$_3$N, which is therefore not abundant in XDRs. The emission of HC$_3$N in the CND must be coming from regions shielded from X-rays. Since the peak of HC$_{3}$N, which may be compact, does not seem to be resolved, the fractional abundance of HC$_{3}$N can only be precisely determined by higher angular resolution data. Therefore, its emission may be coming from hot cores in compact star forming regions, or more cold and yet dense cores ($n \sim$10$^{4}$ cm$^{-3}$) in a more spread out distribution. The existence of HC$_{3}$N suggests that other carbon-chain molecules might also be observable.

\subsubsection{CS and SO}
The model for the CND by Harada et al. (2013) shows that the CS/SO ratio increases with decreasing density both at the free-fall time of a molecular cloud and at steady-state. For both CS and SO, the values in the shocked cloud are more than an order of magnitude higher than those in the cold core. Considering the uncertainty, it is not possible to identify the source of the emission in NGC 1068 by a comparison with these galactic sources. High abundances of the sulfur-containing species SO and CS may result from high elemental sulfur abundance in the gas phase due to shock waves (Wakelam et al. 2005).

\subsubsection{Interpretation of Chemical Features}

Overall, the CND hosts relatively complex molecules, which are often associated with shocked molecular clouds or hot cores. Because a high X-ray flux can dissociate these molecules, they must also reside in regions shielded from X-rays.
 
Although we have derived the column densities of these molecules, there are some difficulties in discerning which exact source the emission is coming from. In this paper, we took the column density ratios over CO column densities. Since the critical density of CO is lower than the rest of the molecules, CO emission can also come from translucent clouds of $n \sim 10^3\,$cm$^{-3}$, which are likely to have a higher volume filling factor. Even if most of the emission of a molecule such as CH$_{3}$CN comes from hot cores, the calculated column density ratio of CH$_{3}$CN over CO will turn out to be lower than the value in Table 7 if the mean density is low. Therefore, it is hard to determine whether the CH$_{3}$CN or HC$_{3}$N emission is coming from an extended volume of molecular clouds with $n \sim 10^4\,$cm$^{-3}$ with a low fractional abundance or from hot cores of $n \sim 10^6\,$cm$^{-3}$ and $T \sim 300\,$K with a small volume filling factor.

Besides the difference in the density, the difference in abundances of each type of source may also explain the observed emission. For example, low abundances of CH$_{3}$CN and HC$_{3}$N are seen in shocked clouds, and their abundances are high in hot cores, so that the differences between the peak locations of CH$_{3}$OH and HNCO and those of CH$_{3}$CN and HC$_{3}$N may come from different environments; viz., hot cores versus shocked clouds. These differences indicate a higher star formation activity within tens of parsecs from the AGN core, where CH$_{3}$CN and HC$_{3}$N have peak abundances, albeit considerably lower than expected from hot core material.

It should be also noted that the shock chemistry might also be different around the AGN. The shocked molecular cloud value used in Table 7 is from an outflow of the young stellar object L1157. Larger scale shocks in the CND of NGC 1068 may be different in both the pre-shock density and evolutionary timescale of the chemistry.

\section{Conclusions}

We observed the Seyfert 2 galaxy NGC 1068 with the 0.8-mm band in the CND and starburst ring during the ALMA early science program. The analyses were carried out with our data obtained at 3-mm band (Paper-1). The main results of this work are summarized in the following list.\\
\begin{enumerate}
\item We successfully observed images of $^{13}$CO ($\it{J}$=3--2), C$^{18}$O ($\it{J}$=3--2), CN ($\it{N}$=3--2), and CS ($\it{J}$=7--6) with an angular resolution of $\sim$1.$\!^{\prime\prime}$3 $\times$ 1.$\!^{\prime\prime}$2. $^{13}$CO and CN were detected in both of the CND and starburst ring, while CS was only detected in the CND. C$^{18}$O was not significantly detected toward the CND, but is relatively strong in the starburst ring.

\item At present, some molecules have been found to be separated into Eastern and Western knots in the CND with interferometric observations, and similar distributions of $^{13}$CO ($\it{J}$=3--2), CN ($\it{N}$=3--2) and CS ($\it{J}$=7--6) are found in our observations for the first time.

\item We have determined rotation diagrams of observed molecules in the 3-mm band (Paper-1) and 0.8-mm band (this work) toward the CND and the southwest position in the starburst ring. The rotational temperatures and column densities have been calculated from these diagrams. $^{13}$CO and C$^{18}$O molecules trace cold gas ($<$ 10 K) in the starburst ring and a little warmer gas (14.2--21.5 K) in the CND. Although the column densities of the CO isotopic species in the CND are only one-third of that in the starburst ring, those of the shock/dust related molecules are enhanced by a factor of 3--4 in the CND under the assumption of the same $T_{\rm{rot}}$ in the CND and starburst ring. If the $T_{\rm{rot}}$ in the CND is higher than the starburst ring, the column density of the shock related molecules, such as HNCO, CH$_{3}$CN, SO, and CH$_{3}$OH in the CND are much larger than that in the starburst ring.

\item The difference of the distributions between $^{13}$CO and C$^{18}$O is unexplained by only a difference of excitation conditions, because the abundance ratio of these species is not sensitive to the molecular gas temperature and density. The column density ratios of $^{13}$CO/C$^{18}$O in the CND and the starburst ring are $\sim$6 under the assumption that $T_{\rm{rot}}$ = 30 K for C$^{18}$O and $\sim$3, respectively. The amount of C$^{18}$O in the CND is smaller than that in the starburst ring as compared with that of $^{13}$CO. This feature is possibly the result of differences in isotopic ratios and/or the effect of isotope-selective photodissociation.

\item We found an especially large enhancement of CN in the CND. Specifically, the column density in the CND is a factor of 20--50 larger than that in the starburst ring. CN is one of the key molecules to study an X-ray dominated region (XDR), as mentioned in theoretical studies (e.g. Meijerink et al. 2007; Harada et al. 2013). Moreover the fractional abundance derived in the CND is closer to the value in XDRs and in the shocked region L1157(B1), but higher than that of cold cores or, especially, hot cores. Since a high X-ray luminosity from the AGN is seen, the high abundance of CN seen in the CND can be explained by XDRs. Therefore, the enhancement of column density and abundance of CN in the CND may be the effects of extreme physical conditions such as those in XDRs.

\item We discuss the chemistry of each species, and compare the observed abundances in the CND and starburst ring with those of Galactic sources such as cold cores, hot cores, and shocked molecular clouds in order to study the overall characteristics. As a result, the CND of NGC 1068 seems to be chemically very rich. The CND hosts relatively complex molecules, which are often associated with shocked molecular clouds or hot cores. Because a high X-ray flux can dissociate these molecules, they must also reside in regions shielded from X-rays.

\end{enumerate}

\bigskip

This paper makes use of the following ALMA data: ADS/JAO.ALMA\#2011.0.00061.S. ALMA is a partnership of ESO (representing its member states), NSF (USA) and NINS (Japan), together with NRC (Canada) and NSC and ASIAA (Taiwan), in cooperation with the Republic of Chile. The Joint ALMA Observatory is operated by ESO, AUI/NRAO and NAOJ. We thank the ALMA staff, especially Akiko Kawamura, for their supports. E. H. wishes to acknowledge the support of the National Science Foundation (US) for his astrochemistry program. 
He also acknowledges support from the NASA Exobiology and Evolutionary Biology program through a subcontract from Rensselaer Polytechnic Institute.

\clearpage

\begin{figure*}
  \begin{center}
    \FigureFile(160mm,160mm){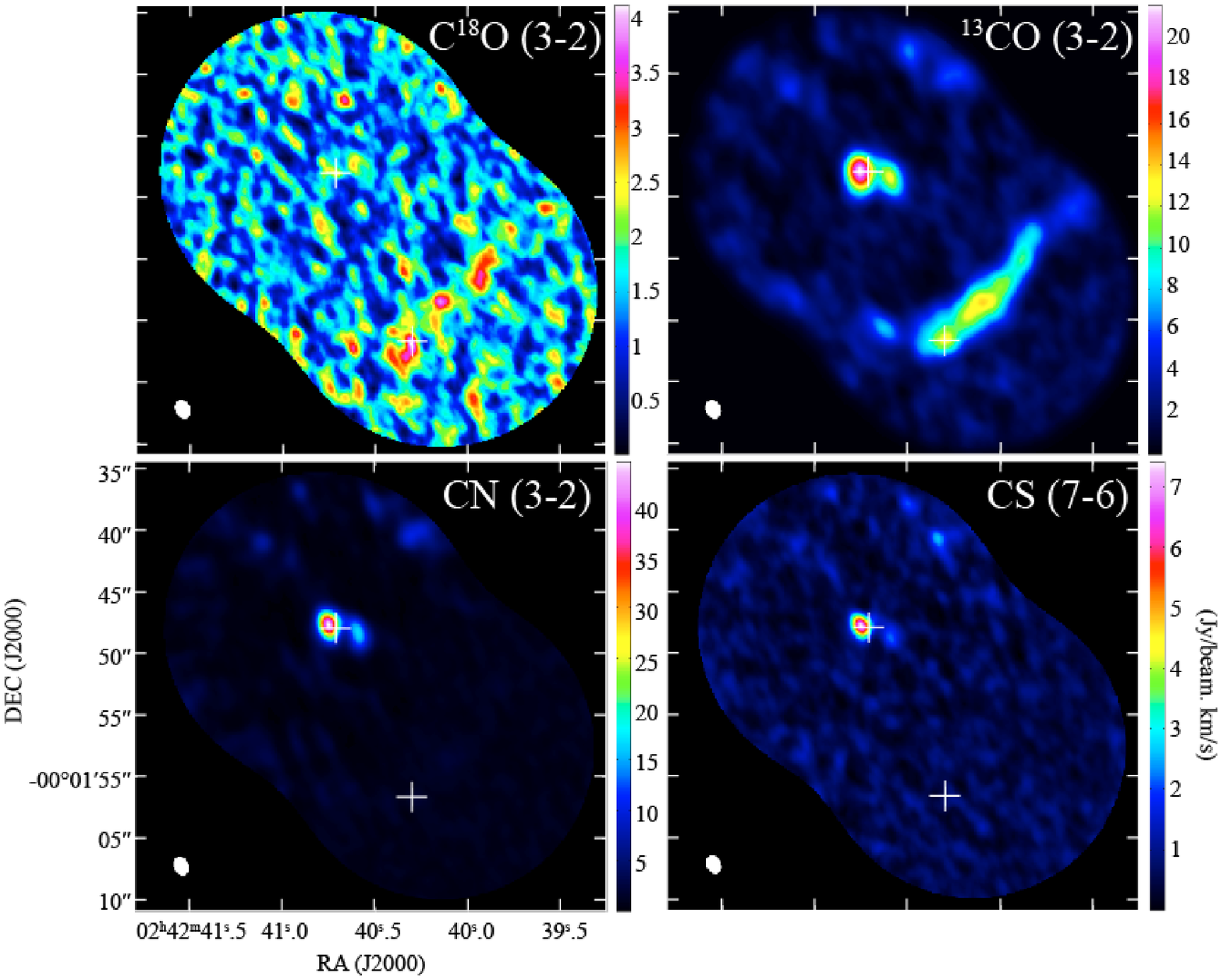}
  \end{center}
  \caption{The images of integrated intensity of C$^{18}$O $\it{J}$ = 3--2, $^{13}$CO $\it{J}$ = 3--2, CN $\it{N}$ = 3--2, and CS $\it{J}$ = 7--6. The central radio continuum position ($\alpha_{J2000}$ = 2$^{h}$42$^{m}$40.$\!^{s}$70912 and $\delta_{J2000}$ =-00$^{\circ}$00$^{\prime}$47.$\!^{\prime\prime}$938; Gallimore et al. 2004) and the $^{13}$CO $\it{J}$ = 3--2 intensity peak at the southwest position in the starburst ring ($\alpha_{J2000}$ = 2$^{h}$42$^{m}$40.$\!^{s}$298 and $\delta_{J2000}$ =-00$^{\circ}$01$^{\prime}$01.$\!^{\prime\prime}$638) are indicated with white crosses. The beam is shown with a white elliptical in the bottom-left corner in each image. The primary beam correction is not applied.}\label{fig1}
\end{figure*}

\clearpage

\begin{figure*}
  \begin{center}
    \FigureFile(160mm,160mm){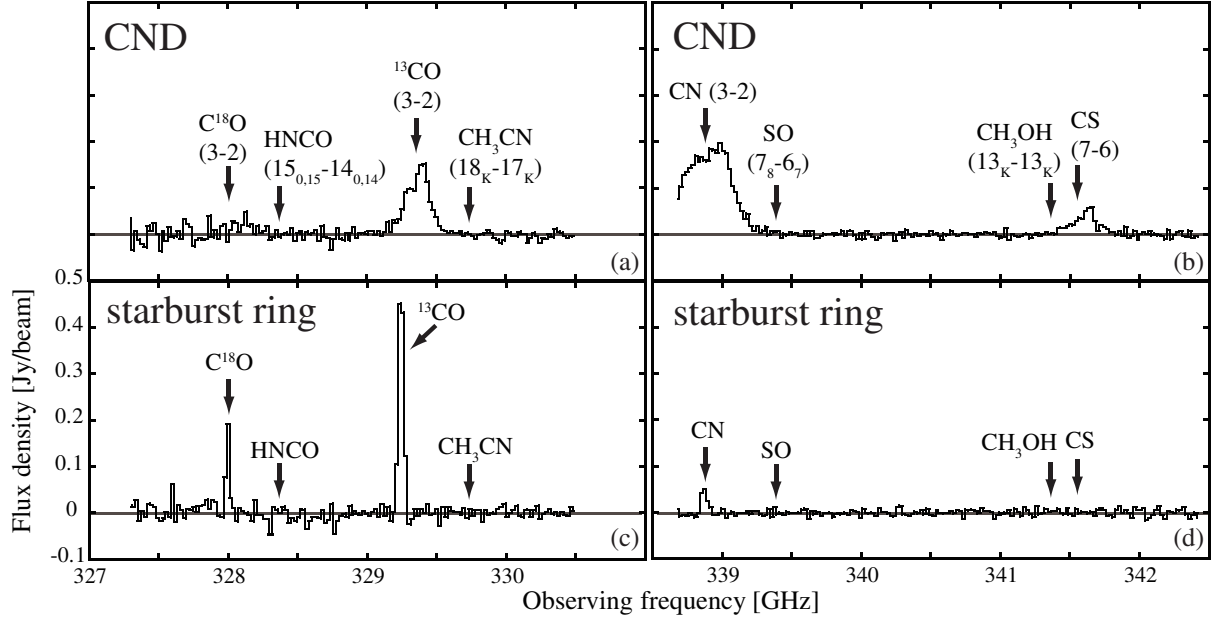}
  \end{center}
  \caption{The top spectra (a) and (b) are at the central continuum position (in the CND), and the bottom spectra (c) and (d) are at the position of the $^{13}$CO $\it{J}$ = 3--2 intensity peak at the southwest position in the starburst ring (see the caption of Figure 1). The primary beam correction and convolution to the beam size of the 3-mm band are applied. The observing frequency is shown as a topocentric value, which is a default reference frame of ALMA. It is necessary to shift the frequency corresponding to $V_{\rm{LSR}}$ = 1150 km/s to obtain the approximate rest frequency. }\label{fig2}
\end{figure*}

\clearpage

\begin{figure*}
  \begin{center}
    \FigureFile(160mm,160mm){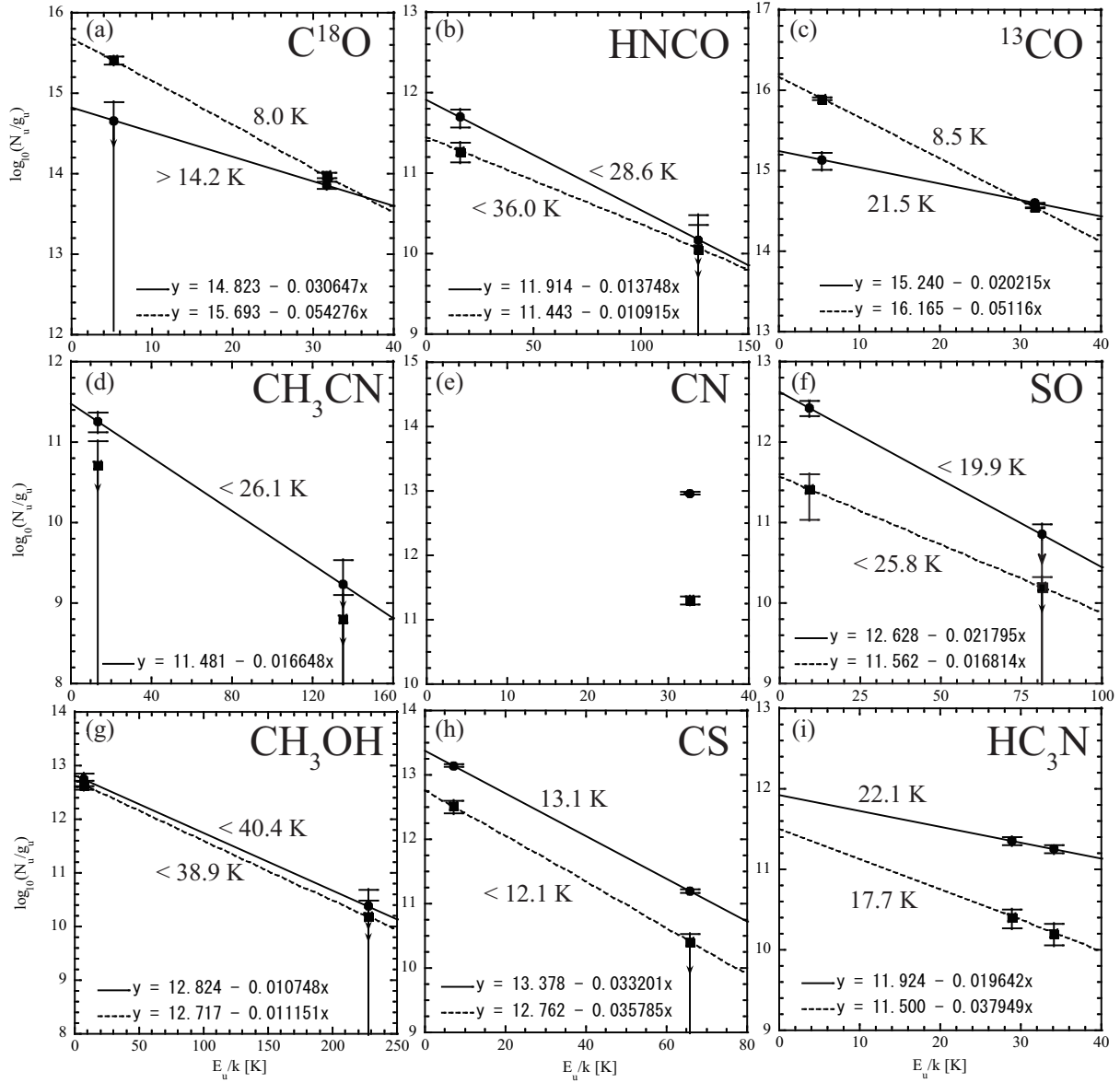}
  \end{center}
  \caption{Rotation diagrams of molecules in the 3-mm band and the 0.8-mm band with ALMA observations. The results of the 3-mm band are listed in Table 2 in Paper-1. The filled circle symbols and solid lines represent results for the CND, and the filled square symbols and dotted lines represent results for the starburst ring.}\label{fig3}
\end{figure*}

\clearpage

\begin{table}
  \caption{Observing log.}\label{tab:log}
  \begin{center}
    \begin{tabular}{lc}
      \hline
      Parameter & Value \\
      \hline
      Observation date & 28 Nov 2011 \\
      Number of antennas & 14 \\
      Observing time (min) & 71.4 \\
      Bandpass calibrator & J0423-013 \\
      Flux calibrator & Callisto \\
      Phase calibrator & J0339-017 \\
      Rms noise (mJy beam$^{-1}$)\footnotemark[$*$] & (CND/starburst ring) \\
      spw0 & 8.6--14.3/9.0--13.6 \\
      spw1 & 5.4--6.6/5.4--7.6 \\
      spw2 & 3.2--4.5/3.3--3.7 \\
      spw3 & 3.5--4.0/3.6--4.5 \\
      Synthesized beam\footnotemark[$\dagger$] & (major $\times$ minor, P.A.) \\
       & 1.$\!^{\prime\prime}$29$\times$1.$\!^{\prime\prime}$23, 2.$\!^{\circ}$3\\
      \hline
      \multicolumn{2}{@{}l@{}}{\hbox to 0pt{\parbox{80mm}{\footnotesize
       \footnotemark[$*$]See the section 2 for details. \footnotemark[$\dagger$]This parameter is for the spectral window of the lowest frequency (spw0).
     }\hss}}
    \end{tabular}
  \end{center}
\end{table}

\begin{table*}
  \caption{Line parameters in the circumnuclear disk (CND).}\label{tab:par}
  \begin{center}
    \begin{tabular}{llllllll}
      \hline
      Frequency\footnotemark[$*$] & Molecule & Transition & $E_{\rm u}$/k & Peak flux & $V_{\rm LSR}$ & Line width & Integrated flux \\
      (GHz) & & & (K) & (Jy beam$^{-1}$) & (km s$^{-1}$) & (km s$^{-1}$) & (Jy beam$^{-1}$ km s$^{-1}$) \\
      \hline
      329.330546 & C$^{18}$O & $\it{J}$ = 3--2 & 31.76 & 0.02$\pm$0.01 & 1106$\pm$20 & 173$\pm$47 & 4.59$\pm$1.16 \\
      329.66437 & HNCO & $\it{J_{Ka,Kc}}$ = 15$_{0,15}$--14$_{0,14}$ & 126.69 & --- & --- & --- & $<$0.97 (1$\sigma$) \\
      330.587960 & $^{13}$CO & $\it{J}$ = 3--2 & 31.64 & 0.13$\pm$0.01 & 1102$\pm$4 & 178$\pm$9 & 25.49$\pm$1.26 \\
      331.071548 & CH$_{3}$CN & $\it{J_{K}}$ = 18$_{K}$--17$_{K}$ & 151.09 & --- & --- & --- & $<$0.83 (1$\sigma$) \\
      340.031567 & CN\footnotemark[$\dagger$] & $\it{N}$ = 3--2, $\it{J}$ = 5/2-3/2 & 32.66 & 0.165$\pm$0.004 & 1049$\pm$8 & 272$\pm$35 & 47.67$\pm$4.38 \\
      340.247874 & CN\footnotemark[$\dagger$] & $\it{N}$ = 3--2, $\it{J}$ = 7/2-5/2 & 32.69 & 0.19$\pm$0.01 & 1171$\pm$7 & 308$\pm$15 & 62.83$\pm$2.52 \\
      340.714350 & SO & $\it{J_{N}}$ = 7$_{8}$--6$_{7}$ & 81.31 & --- & --- & --- & $<$0.76 (1$\sigma$) \\
      342.729781 & CH$_{3}$OH & $\it{J_{Ka,Kc}}$ = 13$_{1,12}$--13$_{0,13}A$-+ & 232.49 & --- & --- & --- & $<$0.49 (1$\sigma$) \\
      342.882866 & CS & $\it{J}$ = 7--6 & 65.88 & 0.046$\pm$0.004 & 1109$\pm$7 & 183$\pm$17 & 8.88$\pm$0.76 \\
      \hline
      \multicolumn{8}{@{}l@{}}{\hbox to 0pt{\parbox{180mm}{\footnotesize
       \footnotemark[$*$] Rest frequency listed by Lovas (2004) except for HNCO. The rest frequency of HNCO is taken from the Cologne Database for Molecular Spectroscopy (CDMS; M$\ddot{\rm u}$ller et al. 2005). We here list frequencies of $\it{J_{K}}$=18$_{0}$--17$_{0}$ for CH$_{3}$CN, $\it{F}$=7/2-5/2, $\it{J}$=5/2-3/2 for CN, and $\it{F}$=9/2-7/2, $\it{J}$=7/2-5/2 for CN. \footnotemark[$\dagger$] Blended lines, see Section 3.2 for details.
     }\hss}}
    \end{tabular}
  \end{center}
\end{table*}

\begin{table*}
  \caption{Line parameters in the starburst ring.}\label{tab:par}
  \begin{center}
    \begin{tabular}{llllllll}
      \hline
      Frequency & Molecule & Transition & $E_{\rm u}$/k & Peak flux & $V_{\rm LSR}$ & Line width & Integrated flux \\
      (GHz) & & & (K) & (Jy beam$^{-1}$) & (km s$^{-1}$) & (km s$^{-1}$) & (Jy beam$^{-1}$ km s$^{-1}$) \\
      \hline
      329.330546 & C$^{18}$O & $\it{J}$ = 3--2 & 31.76 & 0.20$\pm$0.02 & 1216$\pm$1 & 28$\pm$3 & 6.02$\pm$0.50 \\
      329.66437 & HNCO & $\it{J_{Ka,Kc}}$ = 15$_{0,15}$--14$_{0,14}$ & 126.69 & --- & --- & --- & $<$0.75 (1$\sigma$) \\
      330.587960 & $^{13}$CO & $\it{J}$ = 3--2 & 31.64 & 0.53$\pm$0.02 & 1216$\pm$1 & 40$\pm$2 & 22.30$\pm$0.80 \\
      331.071548 & CH$_{3}$CN & $\it{J_{K}}$ = 18$_{K}$--17$_{K}$ & 151.09 & --- & --- & --- & $<$0.41 (1$\sigma$) \\
      340.031567 & CN & $\it{N}$ = 3--2, $\it{J}$ = 5/2-3/2 & 32.66 & --- & --- & --- & $<$0.27 (1$\sigma$) \\
      340.247874 & CN & $\it{N}$ = 3--2, $\it{J}$ = 7/2-5/2 & 32.69 & 0.05$\pm$0.01 & 1212$\pm$3 & 42$\pm$6 & 2.41$\pm$0.34 \\
      340.714350 & SO & $\it{J_{N}}$ = 7$_{8}$--6$_{7}$ & 81.31 & --- & --- & --- & $<$0.27 (1$\sigma$) \\
      342.729781 & CH$_{3}$OH & $\it{J_{Ka,Kc}}$ = 13$_{1,12}$--13$_{0,13}A$-+ & 232.49 & --- & --- & --- & $<$0.31 (1$\sigma$) \\
      342.882866 & CS & $\it{J}$ = 7--6 & 65.88 & --- & --- & --- & $<$0.31 (1$\sigma$) \\
      \hline
      \multicolumn{8}{@{}l@{}}{\hbox to 0pt{\parbox{180mm}{\footnotesize
       See the caption of Table 2 for details.
     }\hss}}
    \end{tabular}
  \end{center}
\end{table*}

\begin{table*}
  \caption{Column densities of C$^{18}$O, $^{13}$CO, CS and HC$_{3}$N.}\label{dens}
  \begin{center}
    \begin{tabular}{lccccc}
      \hline
      & \multicolumn{2}{c}{CND} & & \multicolumn{2}{c}{starburst ring} \\ 
      \cline{2-3}
      \cline{5-6}
      Molecule\footnotemark[$*$] & $T_{\rm{rot}}$ & $N_{\rm{mol}}$ & & $T_{\rm{rot}}$ & $N_{\rm{mol}}$ \\
      & (K) & ($\times$10$^{14}$ cm$^{-2}$) & & (K) & ($\times$10$^{14} $cm$^{-2}$) \\
      \hline
      C$^{18}$O & --- & --- & & 8.0$\pm$0.5 & 167$^{+19}_{-18}$ \\
      $^{13}$CO & 21.5$^{+6.4}_{-3.4}$ & 147$^{+14}_{-10}$  && 8.5$\pm$0.2 & 521$^{+18}_{-19}$ \\
      CS & 13.1$\pm$0.3 & 2.7$^{+0.2}_{-0.1}$ & & --- & --- \\
      HC$_{3}$N & 22 & 0.9 & & 18 & 0.1 \\
      \hline
      \multicolumn{5}{@{}l@{}}{\hbox to 0pt{\parbox{100mm}{\footnotesize
       \footnotemark[$*$] For these molecules, with two transitions available, we calculated $T_{\rm{rot}}$ from rotation diagrams (see also Figure 3). 
     }\hss}}
    \end{tabular}
  \end{center}
\end{table*}

\begin{table*}
  \caption{Column densities ($\times$10$^{14}$ cm$^{-2}$) for the assumed rotation temperatures of the observed molecules.}\label{dens}
  \begin{center}
    \begin{tabular}{lccccccccccc}
      \hline
      & \multicolumn{5}{c}{CND} &  & \multicolumn{5}{c}{starburst ring} \\ 
      \cline{2-6}
      \cline{8-12}
      Molecule\footnotemark[$*$] & 5(K) & 10(K) & 15(K) & 30(K) & 45(K) & & 5(K) & 10(K) & 15(K) & 30(K) & 45(K) \\
      \cline{1-6}
      \cline{8-12}
      C$^{18}$O & 901.9 & 69.8 & 35.4 & 24.0 & 25.1 & & --- & --- & --- & --- & --- \\
      HNCO & 1.2 & 0.5 & 0.4 & 0.5 & 0.6 & & 0.3 & 0.1 & 0.1 & 0.1 & 0.2 \\
      CH$_{3}$CN & 0.2 & 0.08 & 0.07 & 0.09 & 0.1 & & --- & --- & --- & --- & --- \\
      CN & 38.4 & 2.9 & 1.5 & 1.0 & 1.0 & & 0.8 & 0.06 & 0.03 & 0.02 & 0.02 \\
      SO & 1.2 & 1.1 & 1.4 & 2.4 & 3.6 & & 0.2 & 0.2 & 0.2 & 0.4 & 0.6 \\
      CH$_{3}$OH & 2.4 & 2.5 & 3.0 & 5.4 & 8.5 & & 1.8 & 1.9 & 2.3 & 3.5 & 5.7 \\
      CS & --- & --- & --- & --- & --- & & 0.4 & 0.4 & 0.5 & 0.7 & 1.0 \\
      \hline
      \multicolumn{11}{@{}l@{}}{\hbox to 0pt{\parbox{130mm}{\footnotesize
       \footnotemark[$*$] For these molecules, with only one transition available, we assumed $T_{\rm{rot}}$ = 10$\pm$5 K and 30$\pm$15 K. 
     }\hss}}
    \end{tabular}
  \end{center}
\end{table*}

\begin{table*}
  \caption{Individual fluxes of the molecular lines from the two knots in the CND.}\label{tab:log}
  \begin{center}
    \begin{tabular}{lccccc}
      \hline
      Molecular Line & Component & $\Delta\alpha$\footnotemark[$*$] & $\Delta\delta$\footnotemark[$*$] & Integrated Flux & Flux Ratio \\
      & & ($^{\prime\prime}$) & ($^{\prime\prime}$) & (Jy beam$^{-1}$ km s$^{-1}$) & (E-knot/W-knot) \\
      \hline
      $^{13}$CO ($\it{J}$=3--2) & E-knot & +0.7 & 0.0 & 16.8$\pm$0.6 & \\
      & W-knot & -1.8 & +0.5 & 7.7$\pm$0.7 & 2.2$\pm$0.2 \\
      CN ($\it{N}$=3--2) & E-knot & +0.6 & -0.1 & 59.8$\pm$3.3 & \\
      & W-knot & -1.7 & +0.8 & 26.2$\pm$3.8 & 2.3$\pm$0.4 \\
      CS ($\it{J}$=7--6) & E-knot & +0.6 & -0.2 & 7.3$\pm$0.5 & \\
      & W-knot & -1.8 & +0.4 & 2.1$\pm$0.4 & 3.5$\pm$0.7 \\
      \hline
      \multicolumn{6}{@{}l@{}}{\hbox to 0pt{\parbox{140mm}{\footnotesize
       \footnotemark[$*$] The offset from the central radio continuum position ($\alpha_{J2000}$ = 2$^{h}$42$^{m}$40.$\!^{s}$70912 and $\delta_{J2000}$ =-00$^{\circ}$00$^{\prime}$47.$\!^{\prime\prime}$938; Gallimore et al. 2004). 
     }\hss}}
    \end{tabular}
  \end{center}
\end{table*}

\begin{table*}
  \caption{The estimated fractional abundances of molecules in the CND and starburst ring of NGC 1068 along with various Galactic sources as references.}\label{tab:log}
  \begin{center}
    \begin{tabular}{lcccccc}
      \hline
      Species & CND & Starburst ring & Cold Core & \multicolumn{2}{c}{Hot Cores} & Shocked Cloud \\
      & (NGC 1068) & (NGC 1068) & (TMC-1) & (Sgr B2(N)) & (AFGL2591) & (L1157(B1)) \\
      \hline
HNCO & 0.7$^{+1.0}_{-0.1}$(-8) & 0.4$^{+1.2}_{-0.0}$(-9) & 5.7$\pm$0.4(-10)\footnotemark[$*$] & 5.6(-10)(halo)\footnotemark[$\ddagger$] & --- & 1.1$\pm$0.7(-8)\footnotemark[$\dagger\dagger$]\\
 & & & & 1.2(-8)(core)\footnotemark[$\ddagger$] & & \\
CH$_3$CN & 1.1$^{+1.6}_{-0.1}$(-9) & ---  & 6(-10)\footnotemark[$\dagger$] & 2.2(-7)(core)\footnotemark[$\ddagger$] & --- & 0.7$\pm$0.4(-9)\footnotemark[$\ddagger\ddagger$]\\
CN & 0.5$^{+6.3}_{-0.3}$(-7) & 0.2$^{+2.8}_{-0.1}$(-9) & 5(-9)\footnotemark[$\dagger$] & --- &  7.0$\pm$0.7(-12)\footnotemark[$\#$] & 1.6(-7)\footnotemark[$\S\S$]\\
SO & 1.6$^{+0.3}_{-0.1}$(-8) & 7.7$\pm$0.0(-10) & 2(-9)\footnotemark[$\dagger$] & 2(-8)(halo)\footnotemark[$\S$] & 1(-8)\footnotemark[$**$] & 2.5(-7)\footnotemark[$\S\S$]\\
CH$_3$OH & 3.4$^{+0.7}_{-0.1}$(-8) & 7.3$^{+1.5}_{-0.4}$(-9) & 3(-9)\footnotemark[$\dagger$] & 3.4(-9)(halo)\footnotemark[$\ddagger$] & $\leq$2(-8)\footnotemark[$**$] & 1.2$\pm$3.4(-5)\footnotemark[$\S\S$]\\
 & & & & 1.7(-6)(core)\footnotemark[$\ddagger$] & & \\
CS & 3.7$^{+0.3}_{-0.1}$(-8) & 1.5$^{+0.4}_{-0.0}$(-9) & 4(-9)\footnotemark[$\dagger$] & --- & 2(-8)\footnotemark[$**$] & 1.9(-7)\footnotemark[$\S\S$]\\
HC$_3$N & 1.2(-8) & 3.8(-10) & 2(-8)\footnotemark[$\dagger$] & 5(-9)(halo)\footnotemark[$\S$] & 7(-9)\footnotemark[$**$] & 1.0(-8)\footnotemark[$\S\S$]\\
 & & & & 1(-7)(core)\footnotemark[$\|$] & & \\
      \hline
      \multicolumn{7}{@{}l@{}}{\hbox to 0pt{\parbox{165mm}{\footnotesize
The expression a(-b) represents $a\times 10^{-b}$. 

In order to obtain the fractional abundances in the CND and the starburst ring, we assume $T_{\rm rot}$ of 10$\pm$5 K except for CS in the CND and HC$_{3}$N, and adopt $N_{\rm H_2}$ values of $7.4 \times 10^{21}$ cm$^{-2}$ at the CND and $2.6 \times 10^{22}$ cm$^{-2}$ at the starburst ring taken from our $^{13}$CO column densities (see the section 4.2).

\footnotemark[$*$]Marcelino et al. (2009), and $N_{\rm H2}$ adopted are $1.0 \times 10^{22}$ cm$^{-2}$ (Irvine et al. 1991). \footnotemark[$\dagger$]Smith et al. (2004). \footnotemark[$\ddagger$]Garrod et al. (2008). \footnotemark[$\S$]Nummelin et al. (2000). \footnotemark[$\|$]de Vicente et al. (2000). \footnotemark[$\#$]Minh \& Yang (2008), and $N_{\rm H2}$ adopted are $2.0 \times 10^{24}$ cm$^{-2}$ (Jim\'{e}nez-Serra et al. 2012). \footnotemark[$**$]Jim\'{e}nez-Serra et al. (2012). \footnotemark[$\dagger\dagger$]Rodr\'{i}guez-Fern\'{a}ndez et al. (2010). \footnotemark[$\ddagger\ddagger$]Arce et al. (2008). \footnotemark[$\S\S$]Bachiller \& Perez Gutierrez (1997).

     }\hss}}
    \end{tabular}
  \end{center}
\end{table*}

\end{document}